\documentclass{article}

\usepackage{PRIMEarxiv}

\usepackage[utf8]{inputenc} 
\usepackage[T1]{fontenc}    
\usepackage{hyperref}       
\usepackage{url}            
\usepackage{booktabs}       
\usepackage{amsfonts}       
\usepackage{nicefrac}       
\usepackage{microtype}      
\usepackage{lipsum}
\usepackage{fancyhdr}       
\usepackage{graphicx}       
\usepackage{amsmath, bm}
\usepackage{subfig}
\graphicspath{{media/}}     

\title{UniTSA: A Universal Reinforcement Learning Framework for V2X Traffic Signal Control
}

\author{Maonan Wang \thanks{School of Science and Engineering, The Chinese University of Hong Kong, Shenzhen, China and Shanghai AI Laboratory, Shanghai, China.},
Xi Xiong \thanks{Key Laboratory of Road and Traffic Engineering, Ministry of Education, Tongji University, Shanghai, China.},
Yuheng Kan, Chengcheng Xu \thanks{SenseTime Group Limited and Shanghai AI Laboratory, Shanghai, China.},
Man-On Pun \thanks{School of Science and Engineering, The Chinese University of Hong Kong, Shenzhen, Shenzhen 518172, China.}
\thanks{Corresponding author: SimonPun@cuhk.edu.cn.}
}

\begin{document}
\maketitle

\begin{abstract}
Traffic congestion is a persistent problem in urban areas, which calls for the development of effective traffic signal control (TSC) systems. While existing Reinforcement Learning (RL)-based  methods have shown promising performance in optimizing TSC, it is challenging to generalize these methods across intersections of different structures. In this work, a universal RL-based TSC framework is proposed for Vehicle-to-Everything (V2X) environments. The proposed framework introduces a novel agent design that incorporates a junction matrix to characterize intersection states, making the proposed model applicable to diverse intersections. To equip the proposed RL-based framework with enhanced capability of handling various intersection structures, novel traffic state augmentation methods are tailor-made for signal light control systems. Finally, extensive experimental results derived from multiple intersection configurations confirm the effectiveness of the proposed framework. The source code in this work is available at \url{https://github.com/wmn7/Universal_Light}
\end{abstract}


\section{Introduction}
Traffic congestion presents a critical challenge in urban areas worldwide, leading to wasted time for individuals, excessive fuel consumption, and increased greenhouse gas emissions \cite{malik2019evaluation}. To alleviate congestion, conventional traffic signal control (TSC) methods such as fixed-cycle traffic control \cite{miller1963settings}, the Webster method \cite{urbanik2015signal}, and Self-Organizing Traffic Light Control (SOTL) \cite{gershenson2004self} have been developed. However, as cities continue to grow, these traditional traffic management approaches often prove insufficient to handle increasing traffic volumes and dynamic road conditions \cite{tomar2022traffic}. The emergence of V2X technologies in transportation systems offers a promising solution by facilitating communication and data exchange among vehicles, infrastructure, and other road users \cite{tong2019artificial}. By leveraging real-time data derived from vehicles, the traffic management infrastructure can efficiently regulate vehicle and pedestrian movements at intersections by dynamically adapting to the changing traffic conditions \cite{wagner2023spat}.

To control signal lights based on real-time traffic conditions, various RL-based methods have been proposed. These methods employ three primary approaches for adjusting traffic lights, namely ``Choose the next phase", ``Keep or change the phase" and ``Set the phase duration". More specifically, in ``Choose the next phase", the RL agent determines the next phase to activate, allowing for a flexible phase sequence rather than a predetermined one \cite{chin2011exploring, zheng2019learning, chen2020toward, zang2020metalight, liang2022oam, 9392372, zhang2022expression, xiong2019learning, rizzo2019kdd, chu2019multi, oroojlooy2020attendlight, ma2021adaptive, mousavi2017traffic, aslani2019developing, su2023emvlight}. While this approach offers flexibility, it has the potential to confuse drivers since the phase selection may appear random, which could incur increased risks of traffic accidents. For the ``Keep or change the phase" approach, the RL agent decides whether to maintain or change the current phases \cite{van2016coordinated, mannion2016experimental, wei2018intellilight}. Finally, the ``Set the phase duration" approach selects the optimal duration for the current phase from a set of predefined options \cite{xu2013study, aslani2017adaptive, aslani2018traffic}. Through direct interaction with the environment, the RL agent effectively learns to adapt to traffic condition changes based on real-world experiences.

Despite the significant improvements achieved by the aforementioned RL-based methods, a major limitation is that they are designed for interactions of designated structures. In other words, these RL models have to be redesigned and re-trained from scratch when dealing with intersections of different structures including approaching roads, lanes, and phases, which incurs substantial resources in terms of data collection, model development, and testing \cite{Tulay2023road}. Thus, it is crucial to develop universal models that can be easily adapted and deployed across a wide range of intersections. As a result, the implementation of V2X can be efficiently scaled up without requiring extensive customization or redevelopment for each individual intersection \cite{farag2021integration}. In the literature, several generalized models were proposed for different junctions \cite{zang2020metalight, oroojlooy2020attendlight, chen2020toward, liang2022oam}. Despite their good performance, these generalized models are only applicable to those specific intersection configurations considered in their design. Furthermore, these models exhibit performance degradation when encountering intersections of unseen configurations.

Motivated the discussions above, we present a \textbf{Uni}versal \textbf{T}raffic \textbf{S}tate \textbf{A}ugmentation RL-based framework (UniTSA). The proposed framework enables the training of a universal agent using augmented data for TSC. In order to handle intersections of diverse configurations, a junction matrix is developed to describe the intersection state. By using this newly proposed matrix, intersections of different structures can be characterized by matrices of the same size. In addition, the ``Keep or change the current phase'' approach is employed as the action design in this work, ensuring consistent model structures across intersections of different configurations. To cope with unseen intersection structures, five traffic state augmentation methods are developed to enrich the agent's data collection during the training process. These augmentation methods provide more comprehensive training, which ultimately leads to improved performance in intersections of configurations not available in the training set. Furthermore, the Low-Rank Adaptation (LoRA) \cite{hu2021lora} is employed to further fine-tune the model for crucial intersections. Finally, extensive experiments using the Simulation of Urban MObility (SUMO) \cite{lopez2018microscopic} platform are conducted by taking into account intersections of various approaching roads, lanes, and phases. The experimental results demonstrate that the proposed UniTSA model achieves excellent performance even in unseen intersections.

Our contributions can be summarized as follows:
\begin{itemize}
    \item An adaptive TSC framework called UniTSA is proposed for V2X by leveraging a universal RL model built upon novel agent designs that can handle diverse intersection structures. In addition, a fine-tuning mechanism is devised to further enhance the performance in key intersections;
    \item Traffic state augmentation methods are developed for the proposed TSC framework, enhancing the agent's understanding of diverse intersections with improved performance in both training and testing sets;
    \item Extensive experiments on $12$ intersections of diverse structures confirm that the proposed UniTSA model substantially outperforms conventional universal models. Moreover, given a new intersection, UniTSA can fine-tune its pre-trained model to achieve comparable or even better performance with significantly reduced training time as compared to training a new model from scratch.
\end{itemize}

The remainder of the paper is structured as follows: Section~\ref{sec_related_work} provides a summary on the TSC-related work whereas Section~\ref{sec_preliminary} introduces the terminology related to road and traffic signals. After that, Section~\ref{sec_methodology} presents the proposed UniTSA framework and the five traffic state augmentation methods before Section~\ref{sec_experiments} elaborates on the experimental setup and results on UniTSA. Finally, Section~\ref{sec_conclusion} concludes the paper.
	
\section{Related Work} \label{sec_related_work}

Extensive research has been conducted in the field of transportation to study TSC. Conventionally, fixed-time signal control is one of the earliest and widely used approaches in TSC \cite{koonce2008traffic}. These methods rely on pre-determined signal timings based on historical traffic patterns or engineering guidelines. Various optimization techniques have been proposed to determine optimal fixed-time plans for specific intersection configurations, where the Webster \cite{urbanik2015signal} method is one of the most successful TSC methods for the single intersection case. It can calculate the cycle length and phase split for a single intersection according to the traffic volume during a certain period (i.e., past $15$ minutes or $30$ minutes). However, such fixed-time control methods often incur suboptimal performance due to their lack of adaptability to dynamically changing traffic conditions. To cope with this problem, actuated control methods such as Sydney Coordinated Adaptive Traffic System (SCATS) \cite{sims1980sydney}, Max-pressure control \cite{varaiya2013max} and Self-Organizing Traffic Light Control (SOTL) \cite{gershenson2004self} were designed to adaptively adjust signal timings based on real-time traffic demand. Despite their many advantages, these actuated control methods are handicapped by the necessity that expert settings are required for each intersection. Furthermore, the performance of these actuated control methods degrades in complex scenarios.

Recently, RL-based TSC methods have attracted substantial attention due to their outstanding adaptability to real-time traffic conditions and impressive capability of learning the optimal control policies in complex scenarios \cite{wei2019survey}. Generally speaking, these RL-based TSC methods can be categorized into three categories, namely the valued-based methods \cite{chin2011exploring, van2016coordinated, wei2018intellilight, zheng2019learning, chen2020toward, zang2020metalight, liang2022oam, 9392372, zhang2022expression}, the policy-based methods \cite{xiong2019learning, rizzo2019kdd, chu2019multi, oroojlooy2020attendlight, ma2021adaptive} and the actor-critic methods \cite{mousavi2017traffic, aslani2018traffic, aslani2019developing, su2023emvlight}. Despite their good performance, most existing RL-based TSC methods focus on training models for specific intersection configurations or scenarios. A few attempts on training generalized TSC models have been made in the literature. For instance, MetaLight \cite{zang2020metalight} trains a more universal model by incorporating the meta-learning strategy proposed in \cite{hospedales2021meta}. However, MetaLight requires re-training of its model parameters for each new intersection encountered. To overcome this shortcoming, \cite{oroojlooy2020attendlight, chen2020toward, liang2022oam} established universal models through parameter sharing. However, these methods do not preserve the original phase structure of traffic lights. In contrast, our proposed method can maintain the original signal light structure while fine-tuning crucial intersections to achieve significantly improved performance.

\begin{figure}[!htbp]
    \centering
    \includegraphics[width=0.5\linewidth]{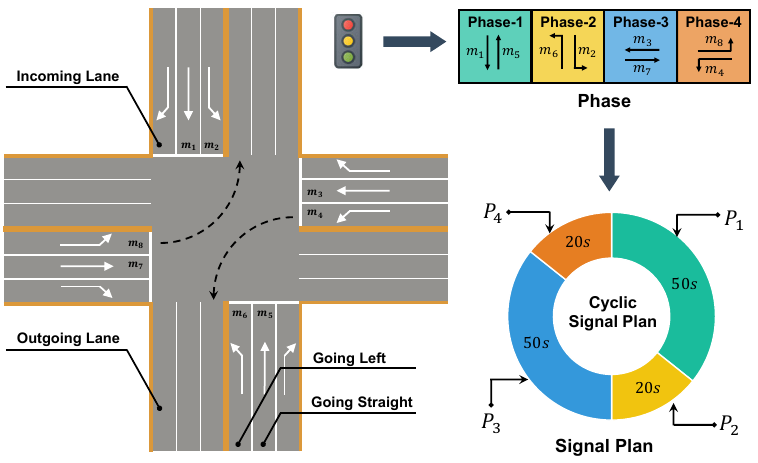}
    \caption{A standard $4$-way intersection illustration.}
    \label{fig_term_defination}
\end{figure}

\section{Preliminary} \label{sec_preliminary}

In this section, some concepts used in this work are defined using a standard four-way intersection as depicted in Fig.~\ref{fig_term_defination} as an example. These concepts can be easily extended to the intersections of different structures. 

\begin{itemize}
    \item Lane: A lane refers to a roadway that provides a defined path for vehicles to travel in a specific direction. At a typical intersection, there are two types of lanes: incoming lanes $l_{in}$ (where vehicles enter) and outgoing lanes $l_{out}$ (where vehicle exit);
    \item Traffic Movement: A traffic movement refers to the connection between an incoming lane $l_{in}$ to an outgoing lane $l_{out}$. For the common $4$-way intersection in the left side of Fig.~\ref{fig_term_defination}, there are $12$ movements in total, including right turns, left turns, and through movements in each of the four directions;
    \item Movement Signal: A movement signal is defined on the traffic movement. The green signal means the corresponding movement is allowed whereas the red signal is prohibited. As the right-turn traffic can move regardless of the signal, only eight movement signals out of the $12$ possible movements in a $4$-way intersection are used. More specifically, these eight movements denoted by $M_{1}, M_{2}, \cdots, M_{8}$ are Northbound (N), Northbound Left-turn (NL), Eastbound (E), Eastbound Left-turn (EL), Westbound(W), Westbound Left-turn (WL), Southbound (S), Southbound Left-turn (SL) movements, respectively. For instance, $m_{8}$ indicates that the vehicles can travel from west to north;
    \item Phase: A phase is a combination of movement signals. Each phase allows a specific group of traffic movements to occur while restricting others. The top-right portion of Fig.~\ref{fig_term_defination} illustrates the four phases of a $4$-way intersection. For instance, Phase-1 involves $m_{1}$ and $m_{5}$, enabling vehicles traveling north-south to proceed while simultaneously prohibiting other movements;
    \item Signal Plan: A signal plan represents a prearranged sequence and duration of phases used to control the traffic signals at an intersection. Mathematically, a signal plan is denoted by $\{(p_{1}, t_{1}), (p_{2}, t_{2}), \cdots, (p_{i}, t_{i}), \cdots \}$, where $p_{i}$ and $t_{i}$ represent a phase and its duration, respectively. Usually, the phase sequence is in a cyclic order. For instance, the bottom-right portion of Fig.~\ref{fig_term_defination} shows a cycle-based signal plan and the duration of each phase is $t_{1}=50$, $t_{2}=20$, $t_{3}=50$ and $t_{4}=20$ as an example. 
\end{itemize}

\begin{figure*}[!htbp]
    \centering
    \includegraphics[width=0.99\textwidth]{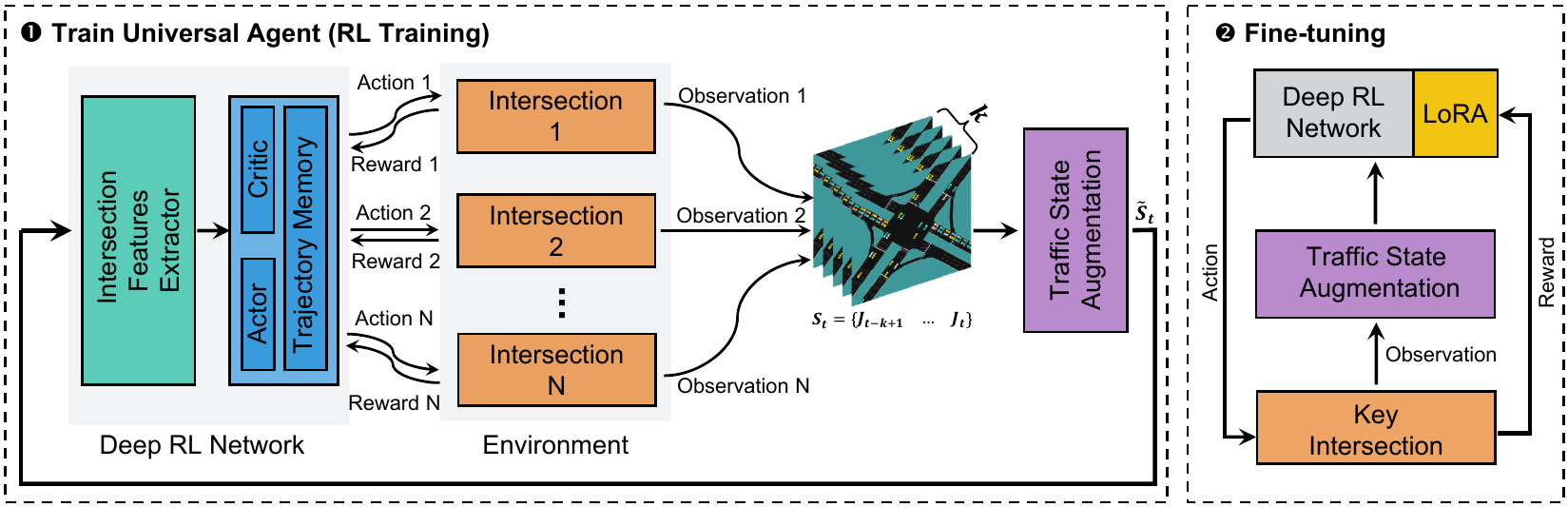}
    \caption{The overall framework of UniTSA.}
    \label{fig_overall_framework}
\end{figure*}

\section{Methodology} \label{sec_methodology}

\subsection{Framework}

As shown in Fig.~\ref{fig_overall_framework}, the proposed UniTSA framework consists of two modules, one designed to train a universal agent through different intersections and one for fine-tuning the model for  key intersections. More specifically, the first module capitalizes on a novel RL agent design to allow the model to characterize intersections of various topologies and signal schemes using the same structure by incorporating features on movements and actions with \textit{next or not} while exploiting five novel traffic state augmentation methods. After that, the second module fine-tunes the model derived by the first module for some specific key intersections. More details about each step above will be elaborated in the following sections.

\subsection{Agent Design and Junction Matrix} \label{sec_agent_design}

\textbf{State}: Intersections may possess different numbers of lanes, which results in state space of different dimensions when recording features in lane units. As discussed in Section~\ref{sec_preliminary}, it is observed that there are only eight valid movement signals in a $4$-way intersection, regardless of the number of lanes. Inspired by this observation, we propose to represent the information of an intersection at time $t$ as a junction matrix denoted as 
\begin{equation}
	{\bm J}_{t} = \left[{\bm m}_{1}^{t}, {\bm m}_{2}^{t}, \cdots, {\bm m}_{8}^{t} \right]^{T},
\end{equation}
where $\left[\cdot\right]^T$ stands for the transpose of the enclosed matrix while vector ${\bm m}_{i}^{t}$ of length eight represents information extracted from the $i$-th movement at time $t$ for $i=1,2,\cdots,8$. More specifically, ${\bm m}_{i}^{t}$ encompasses three components: \textit{traffic characteristics}, \textit{movement characteristics}, and \textit{traffic signal characteristics}. The \textit{traffic characteristics} quantify the congestion level of the movement using parameters such as the average traffic flow $F^{i,t}$, the maximum occupancy $O^{i,t}_{max}$, and the average occupancy $O^{i,t}_{mean}$ within two consecutive actions. In addition, the \textit{movement characteristics} provide specific details about the movement itself, including the direction ($I^{i}_{s}$) indicating whether it is a straight movement or not and the number of lanes ($L_{i}$) it occupies. Finally, the \textit{traffic signal characteristics} comprise three binary parameters, namely $I^{i,t}_{cg}$, $I^{i,t}_{ng}$ and $I^{i,t}_{mg}$. These three parameters indicate whether the movement signal is currently green or not, whether it will be green in the next signal phase or not, and whether the current green duration has reached the minimum duration or not, respectively. These eight features can be readily obtained, making this design suitable for practical deployment. Therefore, vector ${\bm m}^{t}_{i}$ is defined as follows:

\begin{equation} \label{eq_movement_info}
	{\bm m}_{i}^{t} = \left[F^{i,t}, O^{i,t}_{max}, O^{i,t}_{mean}, I^{i}_{s}, L_{i}, I^{i,t}_{cg}, I^{i,t}_{ng}, I^{i,t}_{mg} \right]^T.
\end{equation}

\begin{figure}[!t]
    \centering
    \subfloat[]{\includegraphics[width=0.4\linewidth]{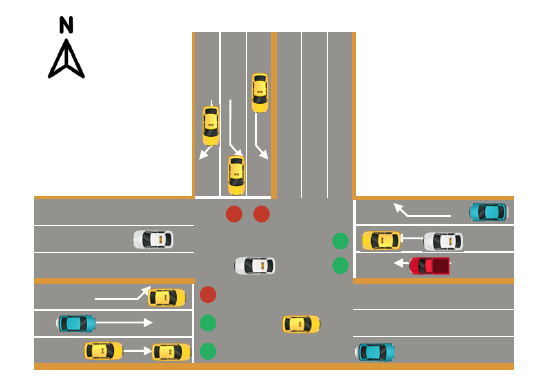}
    \label{fig_threeway_junction}}
    \subfloat[]{\includegraphics[width=0.4\linewidth]{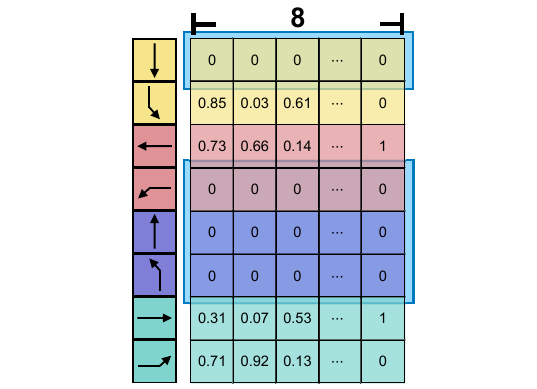}%
    \label{fig_junction_matrix}}
    \caption{
        (a) A $3$-way intersection with (b) its junction matrix with zero padding. 
    }
\label{fig_state}
\end{figure}

It has been observed that ${\bm J}_{t}$ at one time instance is not sufficient to provide a comprehensive understanding of the traffic dynamics for RL-based signal light controllers. To circumvent this obstacle, we propose to incorporate temporal traffic information by exploiting the $K$ latest observations, which enables the agent to better capture traffic patterns and trends in traffic behaviors. As a result, the agent can more effectively adapt its decision-making process in response to evolving traffic conditions. Mathematically, we define the state ${\bm S}_{t} \in \mathbb{R}^{K \times 8 \times 8}$ of the proposed agent at time $t$ as shown in Eq.~\eqref{eq_state}:

\begin{equation} \label{eq_state}
    {\bm S}_{t} = \left[{\bm J}_{t-K+1}, {\bm J}_{t-K+2}, \cdots,  {\bm J}_{t} \right].
\end{equation}

Finally, it is worth pointing out that zero padding is employed when the number of movements at an intersection is less than $8$, e.g. a $3$-way intersection. For instance, Fig.~\ref{fig_threeway_junction} shows a common $3$-way intersection in which only E, EL, W, and SL movement signals are in use. As shown in Fig.~\ref{fig_junction_matrix}, zero padding is applied on the rows corresponding to those $4$ unused movement signals, maintaining the matrix size identical to those of the $4$-way intersection.

\textbf{Action}: A realistic and implementable action design has to take into account the safety of all traffic participants. While the action design \textit{choose next phase} \cite{chen2020toward, oroojlooy2020attendlight, liang2022oam} can significantly improve the intersection efficiency, it disrupts the original cycle sequence of signal lights, thereby compromising driver safety. In sharp contrast, this work adopts the action design of "Keep or change" \cite{van2016coordinated, wei2018intellilight, zheng2019learning}. This design adheres to the concept of cycles, executing each phase sequentially (e.g., Phase 1, Phase 2, Phase 3, Phase 4, Phase 1, Phase 2, and so on). The agent determines whether to keep the current phase or change to the next phase based on the state ${\bm S}_{t}$. Due to the availability of current phase information $I_{cg}$ and next phase information $I_{ng}$ in the junction matrix, it is later shown that this action design exhibits excellent scalability for intersections with different signal plans.

\textbf{Reward}: The negative of the average queue length in each movement $q_{i}$ is adopted as the reward. Metrics such as waiting time, travel time, and delay are not used since it is impractical to obtain these metrics from real-world traffic detection devices. Consequently, the proposed reward function is defined as follows: 

\begin{equation} \label{eq_reward}
    r_{t} = \frac{\left(-\displaystyle\sum_{i=1}^{8}{q_{i}}\right) - \mu}{\sigma + \epsilon},
\end{equation}
where $\epsilon$ is a small number to prevent division by zero. Furthermore, $\mu$ and $\sigma$ represent the mean and standard deviation of the first $R$ rewards, respectively. Mathematically, $\mu$ and $\sigma$ take the following form:
\begin{eqnarray}
\mu &=& \frac{1}{R-1}\displaystyle\sum_{j=1}^{R-1}{r_{j}},\\
\sigma &=& \sqrt{\frac{1}{R-1} \displaystyle\sum_{j=1}^{R-1}{(r_{j} - \mu)^{2}}}.
\end{eqnarray}

The reward is normalized to facilitate a faster training process.

\begin{figure*}[!htbp]
    \centering
    \includegraphics[width=0.95\linewidth]{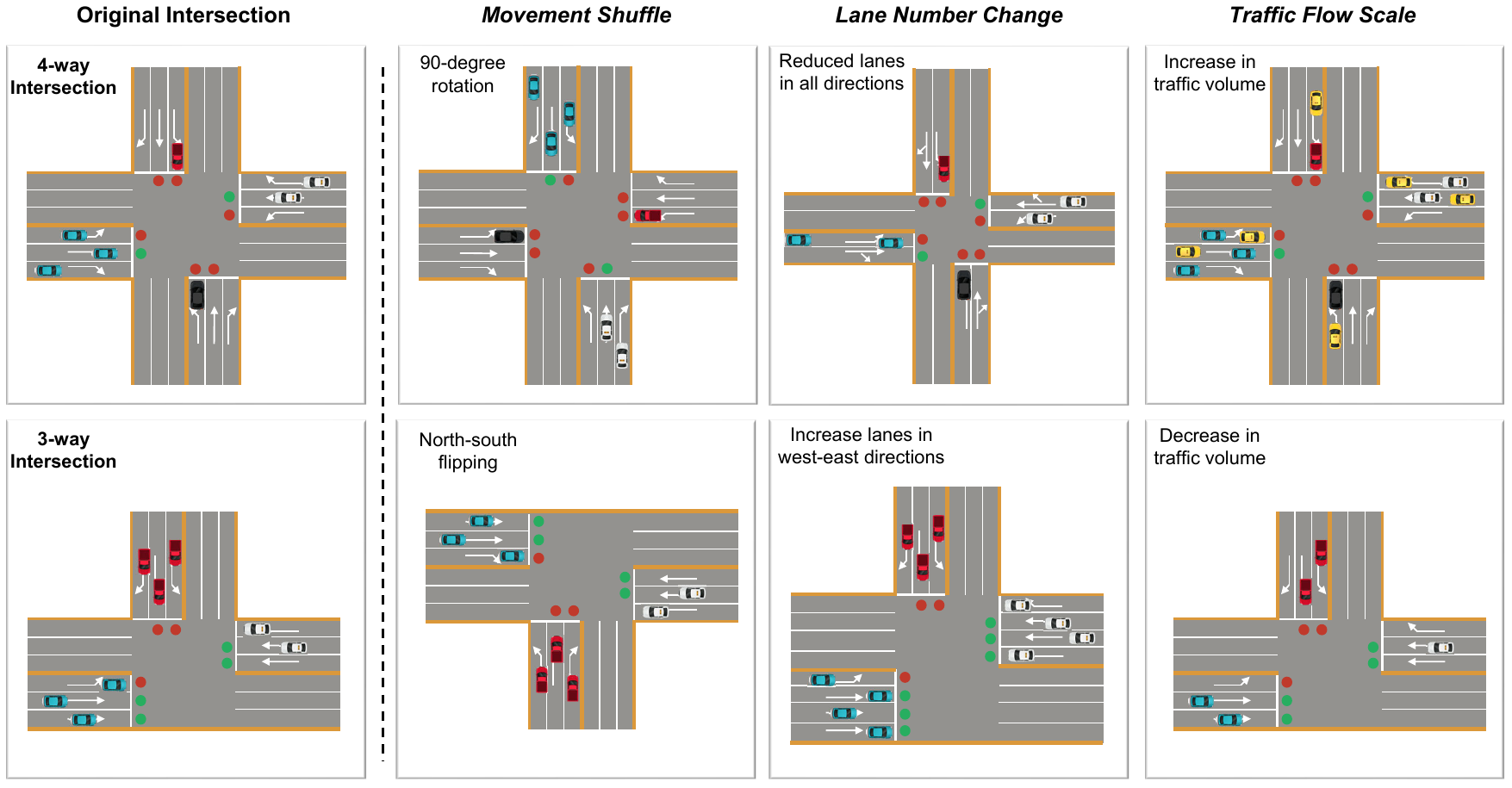}
    \caption{Illustration of three traffic state augmentation methods  applied to both $4$-way and $3$-way intersections.}
    \label{fig_data_augmentation}
\end{figure*}

\begin{figure*}[!htbp]
    \centering
    \includegraphics[width=0.95\linewidth]{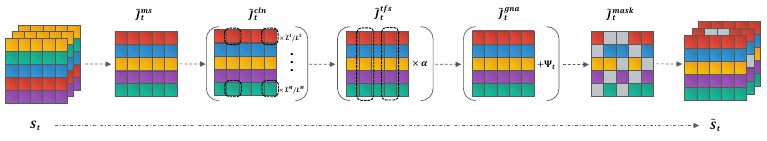}
    \caption{Detailed steps in traffic state augmentation block.}
    \label{fig_data_augmentation_steps}
\end{figure*}

\subsection{Traffic State Augmentation} \label{sec_data_augmentation}

In recent studies, data augmentation techniques have demonstrated their effectiveness in enhancing the generalization capabilities of RL models \cite{laskin2020reinforcement, kostrikov2020image, hansen2021generalization}. By training on a more diverse set of augmented samples, RL agents can improve their capability of handling unseen tasks. The most common data augmentation techniques reported in the literature are \textit{Gaussian noise addition}, and \textit{masking}. In this work, we propose three additional novel traffic state augmentation methods specifically designed for RL-based TSC tasks, namely \textit{movement shuffling}, \textit{change of lane numbers} and \textit{traffic flow scaling} as illustrated in Fig.~\ref{fig_data_augmentation}. 

Fig.~\ref{fig_data_augmentation_steps} provides a detailed illustration of the step-by-step process involving the traffic state ${\bm S}_{t}$ as it undergoes the five data augmentation methods, leading to the final output $\tilde{\bm S}_{t}$. During training, a minibatch of data is randomly sampled from the replay buffer or recently augmented trajectories. While augmentation across the minibatch is stochastic, it is consistent across the stacked frames. It is also worth noting that these five traffic state augmentation methods can be directly applied to the junction matrix ${\bm J}$, enabling the agent to learn and adapt to different traffic scenarios and intersection structures.

\textbf{Movement Shuffling}: This method involves shuffling the rows of the junction matrix, simulating different rotations, and flipping at the same intersection. Intuitively, shuffling can be interpreted as an effective rotation of the original intersection as depicted in Fig.~\ref{fig_data_augmentation}. The assumption behind shuffling is that the action taken by the agent should not change after the rotation of the intersection. Mathematically, the movement shuffling operation can be modeled as follows: 

\begin{equation} \label{eq_movement_shuffle}
    \tilde{\bm J}_{t}^\textrm{ms} = {\bm P} \cdot {\bm J}_{t}, 
\end{equation}
where ${\bm J}_{t}$ and $\tilde{\bm J}_{t}^\textrm{ms}$ represent the original and augmented junction matrices, respectively. Furthermore, ${\bm P}$ is a permutation matrix designed to exchange rows in ${\bm J}_{t}$. After ``movement shuffling'' is applied to all junction matrices in ${\bm S}_{t}$ in Eq.~\eqref{eq_state}, the new state can be represented as $\tilde{\bm S}_{t}^\textrm{ms}$:

\begin{equation} \label{eq_movement_shuffle_state}
    \tilde{\bm S}_{t}^\textrm{ms} = \left[\tilde{\bm J}_{t-K+1}^\textrm{ms}, \tilde{\bm J}_{t-K+2}^\textrm{ms}, \cdots,  \tilde{\bm J}_{t}^\textrm{ms} \right].
\end{equation}

\textbf{Change of Lane Numbers}: To expose the agent to a wider range of road structure combinations, we propose to randomly modify the number of lanes $L_{i}$ in each movement vector $\tilde{\bm m}_{i}^\textrm{ms} \in \tilde{\bm J}_{t}^\textrm{ms}$. This augmentation method allows the agent to encounter various lane configurations during training, enhancing its capability of handling diverse intersection layouts. In addition, the traffic characteristics (e.g., traffic flow and occupancy) are multiplied by the corresponding coefficients to maintain relative values. 

An example of this traffic state augmentation method is shown in Fig.~\ref{fig_data_augmentation}. In this example, a $4$-way intersection was modified to two in all directions by reducing the number of lanes. Similarly, for the $3$-way intersection, we increase the number of west and east-approaching lanes to $4$. While modifying the number of lanes, we maintain the relative number of vehicles in each direction. Thus, it is reasonable to assume that the actions taken by the agent before and after such modifications should be identical. Mathematically, the operation to change the lane number can be expressed as: 

\begin{equation} \label{eq_lane_number_change}
    \tilde{\bm m}_{i}^\textrm{cln} = f(\tilde{\bm m}_{i}^\textrm{ms}, \tilde{L}_{i}), \ i=1,2,\cdots,8,
\end{equation}
where $L_{i}$ and $\tilde{L}_{i}$ are uniformly distributed random variables representing the original and modified numbers of lanes, respectively. Furthermore, $\tilde{\bm m}_{i}^\textrm{ms}$ stands for an individual movement vector within the junction matrix after the ``movement shuffling'' method whereas $f\left(\cdot,\cdot\right)$ denotes the function that adjusts the traffic characteristics. Specifically, $f \left( \cdot, \cdot \right)$ can take the following form:

\begin{equation} \label{eq_lane_number_change_f}
    f(\tilde{\bm m}_{i}^\textrm{ms}, L^{\prime}) = 
    \left\{
        \begin{aligned}
            & \left[\tilde{\bm m}_{i}^\textrm{ms}\right]_{k} \times \left(\frac{\tilde{L}_{i}}{L_{i}}\right), & \text{if}\ k=1,2,3,5 \\
            & \left[\tilde{\bm m}_{i}^\textrm{ms}\right]_{k}, & \text{otherwise}
        \end{aligned}
    \right.
\end{equation}
where $\left[\tilde{\bm m}_{i}^\textrm{ms}\right]_{k}$ represents the $k$-th entry of $\tilde{\bm m}_{i}^\textrm{ms}$. Note that $f$ is designed to ensure that the traffic characteristics $F$, $O_{max}$, and $O_{mean}$ maintain relative values regardless of the variations in the lane configuration. After applying the ``change of lane numbers'' method in all junction matrices in $\tilde{\bm S}_{t}^\textrm{ms}$, the new state can be represented as $\tilde{\bm S}_{t}^\textrm{cln}$:

\begin{equation} \label{eq_lane_number_change_state}
    \tilde{\bm S}_{t}^\textrm{cln} = \left[\tilde{\bm J}_{t-K+1}^\textrm{cln}, \tilde{\bm J}_{t-K+2}^\textrm{cln}, \cdots,  \tilde{\bm J}_{t}^\textrm{cln} \right].
\end{equation}
where 
\begin{equation}\label{eq:Jcln}
\tilde{\bm J}_{t}^\textrm{cln} = \left[\tilde{\bm m}_{1}^\textrm{cln},\tilde{\bm m}_{2}^\textrm{cln}, \cdots, \tilde{\bm m}_{8}^\textrm{cln} \right]^{T}.
\end{equation} 

\textbf{Traffic Flow Scaling}: To shift the agent's focus from absolute car numbers to relative car distributions, we introduce a flow scaling factor. By multiplying the flow and occupancy values in the junction matrix with a uniformly distributed random number $\alpha$, we can create variations in the relative traffic volume for each movement. Notably, $\alpha$ remains consistent across all movements within the same traffic state, ensuring that the relative vehicle proportions between movements remain unchanged. This augmentation method facilitates the agent to prioritize the relative significance of different movements, thereby reducing its reliance on absolute values. It is worth noting that traffic flow scaling does not alter the number of lanes in each movement.

Fig.~\ref{fig_data_augmentation} illustrates two plausible scaling methods by proportionally increasing or decreasing the number of vehicles on each approach or adding small changes to the traffic volume in two scenarios. Mathematically, the flow scaling operation can be defined as: 

\begin{equation} \label{eq_flow_scale}
    \tilde{\bm m}_{i}^\textrm{tfs} = g(\tilde{\bm m}_{i}^\textrm{cln}, \alpha), \ i=1,2,\cdots,8,
\end{equation}
where $\alpha$ is the flow scaling factor. Furthermore, $g\left(\cdot,\cdot\right)$ denotes a function that scales the flow and occupancy values and takes the following form:

\begin{equation} \label{eq_flow_scale_g}
    g(\tilde{\bm m}_{i}^\textrm{cln}, \alpha) = 
    \left\{
        \begin{aligned}
            &\left[\tilde{\bm m}_{i}^\textrm{cln}\right]_{k} \times \alpha, & \text{if}\ k=1,2,3 \\
            & \left[\tilde{\bm m}_{i}^\textrm{cln}\right]_{k}. & \text{otherwise}
        \end{aligned}
    \right.
\end{equation}

After replacing $\tilde{\bm m}_{1}^\textrm{cln}$ in Eq.~\eqref{eq:Jcln} with $\tilde{\bm m}_{i}^\textrm{tfs}$, the resulting state from the ``traffic flow scaling'' method is denoted as $\tilde{\bm S}_{t}^\textrm{tfs}$. 

\textbf{Gaussian Noise Addition}: Gaussian noise is added directly to the junction matrix to introduce randomness into the training data. The additive noise can affect all components of the junction matrix, including traffic characteristics, movement characteristics, and traffic signal characteristics. This augmentation method allows the agent to adapt to noisy and uncertain traffic conditions, improving its robustness during inference. Mathematically, the Gaussian noise addition operation can be modeled as: 

\begin{equation} \label{eq_gaussian_noise}
    \tilde{\bm J}_{t}^\textrm{gna} = \tilde{\bm J}_{t}^\textrm{tfs} + {\bm \Psi}_{t},
\end{equation}
where ${\bm \Psi}_{t} \sim \mathcal N({\bm 0}, {\bm I})$ denotes the Gaussian noise matrix, and $\tilde{\bm J}_{t}^\textrm{tfs} \in \tilde{\bm S}_{t}^\textrm{tfs}$ corresponds to the junction matrix after applying the ``traffic flow scaling'' method. After applying ``Gaussian noise addition'', the new state can be represented as $\tilde{\bm S}_{t}^\textrm{gna}$. 

\textbf{Masking}: To encourage the agent to learn the traffic flow changes, we randomly set the values in the junction matrix to zero at a specific time instance. By masking certain components of the junction matrix, we create situations where the agent must rely on the information before and after the masked period to infer traffic dynamics. This augmentation method promotes the agent's capability of understanding and responding to traffic fluctuations. 

Finally, a novel state denoted as $\tilde{\bm S}_{t}$ is obtained after the application of the five data augmentation methods discussed above.

\begin{figure}[!htbp]
    \centering
    \includegraphics[width=0.6\linewidth]{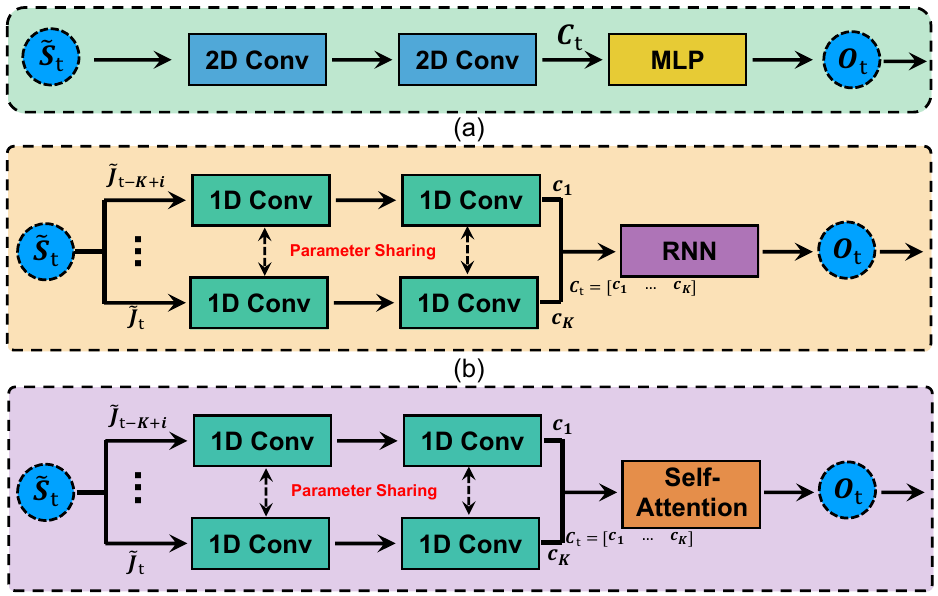}
    \caption{Three kinds of intersection feature extraction block (a) CNN-based Structure; (b) RNN-based Structure; (c) Transformer-based Structure.}
    \label{fig_intersection_feature_extraction}
\end{figure}

\subsection{Intersection Feature Extraction} \label{sec_feature_extraction}

In this section, three neural network structures are utilized to extract intersection information from the augmented traffic states, namely Convolutional Neural Network (CNN), Recurrent Neural Network (RNN) and transformers.

\textbf{CNN-based Structure}: As shown in Fig.~\ref{fig_intersection_feature_extraction}(a), a CNN-based structure equipped with two 2D convolutional layers is utilized to extract time series information within a road junction, resulting in a hidden representation ${\bm C}_{t}$ given as:
\begin{equation} \label{eq_feature_cnn_1}
    {\bm C}_{t} = \text{ReLU}\left(\mathbf{W}^{2d}_{2} \mathbf{W}^{2d}_{1} {\tilde{\bm S}_{t}}\right), 
\end{equation}
where $\mathbf{W}^{2d}_{1}$ and $\mathbf{W}^{2d}_{2}$ represent the learnable parameters of the two 2D convolutional blocks, respectively. More specifically, the first convolutional block, $\mathbf{W}^{2d}_{1}$, extracts information regarding the traffic movement, and subsequently, the second convolutional block, $\mathbf{W}^{2d}_{2}$, captures information specific to the junction based on the movement information. Furthermore, $\text{ReLU}(\cdot)$ denotes the ReLU function. 

Next, the resulting ${\bm C}_{t}$ is passed through a multilayer perceptron (MLP) layers, producing a feature vector ${\bm O}_{t}$ as follows:

\begin{equation} \label{eq_feature_cnn_2}
    {\bm O}_{t} = \mathbf{W}_{u} {\bm C}_{t} + \mathbf{b}_{u}, 
\end{equation}
where $\mathbf{W}_{u}$ and $\mathbf{b}_{u}$ are learnable parameters of the MLP layer. 

\textbf{RNN-based Structure}: In contrast to the CNN-based approach, the RNN-based structure illustrated in Fig.~\ref{fig_intersection_feature_extraction}(b) employs a parameter-sharing 1D convolutional layer to extract information from each junction matrix. The 1D convolutional layer operates on each junction matrix $\tilde{\bm J}_{t-K+i}$ for $i=1,2,\cdots, K$, which is the augmented state at a particular time step within the history window. The resulting outputs denoted by $\bm{c}_1, \bm{c}_2, \ldots, \bm{c}_k$, capture the information from the intersection at $K$ different time instances and can be expressed as:
\begin{equation} \label{eq_feature_rnn_1}
    \bm{c}_{i} = \text{ReLU}\left(\mathbf{W}^{1d}_{2} \mathbf{W}^{1d}_{1} \tilde{\bm J}_{t-K+i}\right), \ i=1,2,\cdots,K,
\end{equation}
where $\mathbf{W}^{1d}_{1}$ and $\mathbf{W}^{1d}_{2}$ represent the learnable parameters of the two 1D convolutional layers. These outputs are then fed into the RNN module: 
\begin{equation} \label{eq_feature_rnn_2}
    \bm{h}_{i} = \tanh(\mathbf{W}_{x} \bm{c}_{i} + \bm{h}_{i-1} \mathbf{W}_{h} + \mathbf{b}_{h}), \ i=1,2,\cdots,K,
\end{equation}
where $\mathbf{W}_x$, $\mathbf{W}_h$, and $\mathbf{b}_h$ represent the weights for the hidden layer in the RNN structure. The final hidden state $\bm{h}_{K}$ derived from the last RNN output is used to calculate the features of the entire intersection over a period of time, denoted as ${\bm O}_{t}$:
\begin{equation} \label{eq_feature_rnn_3}
    {\bm O}_{t} = \mathbf{W}_{v} \bm{h}_{K} + \mathbf{b}_{v}, 
\end{equation}
where $\mathbf{W}_{v}$ and $\mathbf{b}_{v}$ are the weights for the output layer in the RNN structure. 

\textbf{Transformer-based Structure}: Fig.~\ref{fig_intersection_feature_extraction}(c) shows the transformer-based approach. we employ a weight-shared CNN network to extract features from the junction matrix at each time step, as mentioned in Eq.~\eqref{eq_feature_rnn_1}. However, instead of using an RNN block, we utilize a transformer encoder to capture temporal dependencies in the sequence of features $\tilde{\bm{C}}_t = [\bm{c}_1, \bm{c}_2, \ldots, \bm{c}_K]$. To incorporate timing information, we insert a learnable embedding denoted as $\bm{c}_{\text{class}}$ to $\tilde{\bm{C}}_t$ as follows:

\begin{equation} \label{eq_feature_transformer_c}
    \bm{C}_t =\left[\bm{c}_{\text{class}}, \tilde{\bm{C}}_t\right]= \left[\bm{c}_{\text{class}}, \bm{c}_1, \bm{c}_2, \ldots, \bm{c}_K\right].
\end{equation}

The output state of the transformer encoder, based on this modified input sequence, serves as the traffic state representation $\bm{O}_{t}$. In the transformer encoder block, self-attention is calculated as follows:

\begin{equation} \label{eq_feature_transformer_3}
    \bm{Q}_{C} = \bm{C}_{t} \bm{W}_{Q}, \bm{K}_{C} = \bm{C}_{t} \bm{W}_{K}, \bm{V}_{C} = \bm{C}_{t} \bm{W}_{V}, 
\end{equation}

and 

\begin{equation} \label{eq_feature_transformer_2}
    \bm{Z}_{t} = \phi \left( \frac{\bm{Q}_{C} \bm{K}_{C}^{T}}{\sqrt{d}} \right) \bm{V}_{C}, 
\end{equation}
where $\bm{Q}_{C}$, $\bm{K}_{C}$, and $\bm{V}_{V}$ represent the projected query, key and value features, respectively, while $\bm{W}_{Q}$, $\bm{W}_{K}$, and $\bm{W}_{V}$ the corresponding parameter metrics. In addition, $\bm{Z}_{t}$ represents the output of self-attention, and $\phi(\cdot)$ denotes the softmax function. The final output $\bm{O}_{t}$ is the first value of $\bm{Z}_{t}$.
 
\subsection{RL Training and Fine-tuning} \label{sec_train_finetune}

The Proximal Policy Optimization (PPO) algorithm~\cite{schulman2017proximal} is adopted to train our UniTSA model. As depicted in Fig.~\ref{fig_overall_framework}, the agent gathers trajectories consisting of observations, actions, and rewards during the interactions with diverse traffic scenarios of different structures. These trajectories serve as the basis for computing the policy loss and value loss utilized to update the weights of the Actor and Critic networks.

The policy loss quantifies the difference between the current policy and the updated policy derived from the collected trajectories. It encourages the agent to increase the probabilities of actions that lead to higher rewards while reducing the probabilities of actions that lead to lower rewards. Mathematically, the policy loss can be formulated as:

\begin{equation} \label{eq_policy_loss}
    \mathcal{L}_{\text{pf}}(\theta) = \hat{\mathbb{E}}_t \left[ \min \left( r_{t} A_{t}, \text{clip}\left(r_{t}, 1 - \epsilon, 1 + \epsilon \right) A_{t} \right) \right],
\end{equation}
where $\hat{\mathbb{E}}_t$ stands for the expectation operator and
$r_{t}$ represents the ratio between the new policy $\pi_\theta(a_t|s_t)$ and the old policy $\pi_{\theta_{\text{old}}}(a_t|s_t)$ :
\begin{equation}
	r_{t} = \frac{\pi_\theta(a_t|s_t)}{\pi_{\theta_{\text{old}}}(a_t|s_t)}.
\end{equation}

Furthermore, $A_{t} = r_{t} + \gamma V(s_{t+1}) - V(s_{t})$ denotes the advantage function whereas the clip function can ensure stable policy updates. 

The value loss measures the discrepancy between the estimated value function and the actual rewards obtained during the interaction. It makes the value function to better approximate the expected cumulative rewards. The value loss can be defined as:

\begin{equation} \label{eq_value_loss}
    \mathcal{L}_{\text{vf}}(\theta) = \hat{\mathbb{E}}_t \left[ \left( V_\theta(s_t) - \hat{R}_{t} \right)^2 \right],
\end{equation}
where $V_\theta(s_t)$ is the estimated value function under policy $\theta$, and $\hat{R}_{t} = \sum_{k=0}^{\infty}{\gamma^{k} r_{t+k}}$ denotes the reward-to-go. Once the policy loss and the value loss have been calculated, the final objective function is expressed as Eq.~\eqref{eq_rl_loss}:

\begin{equation} \label{eq_rl_loss}
    \mathcal{L}(\theta) = -\mathcal{L}_{\text{pf}}(\theta) + \lambda \mathcal{L}_{\text{vf}}(\theta), 
\end{equation}
where $\lambda$ is the coefficient of value loss.

This work proposes an effective universal model to control the traffic signals by optimizing this objective function through PPO. Furthermore, since certain intersections are more critical than others in practice, LoRA \cite{hu2021lora} is adopted in the proposed model to fine-tune the performance on these important intersections. More specifically,  the LoRA modules are added to the weights of dense layers in the Actor and Critic networks as shown in Fig.~\ref{fig_overall_framework}. During the fine-tuning process, the original pretrained weights are kept constant while the LoRA modules is being updated. This design allows the model to adapt to the intersection-specific features without significantly increasing the number of parameters, while maintaining training efficiency. 

\begin{figure*}[!htbp]
    \centering
    \includegraphics[width=0.99\linewidth]{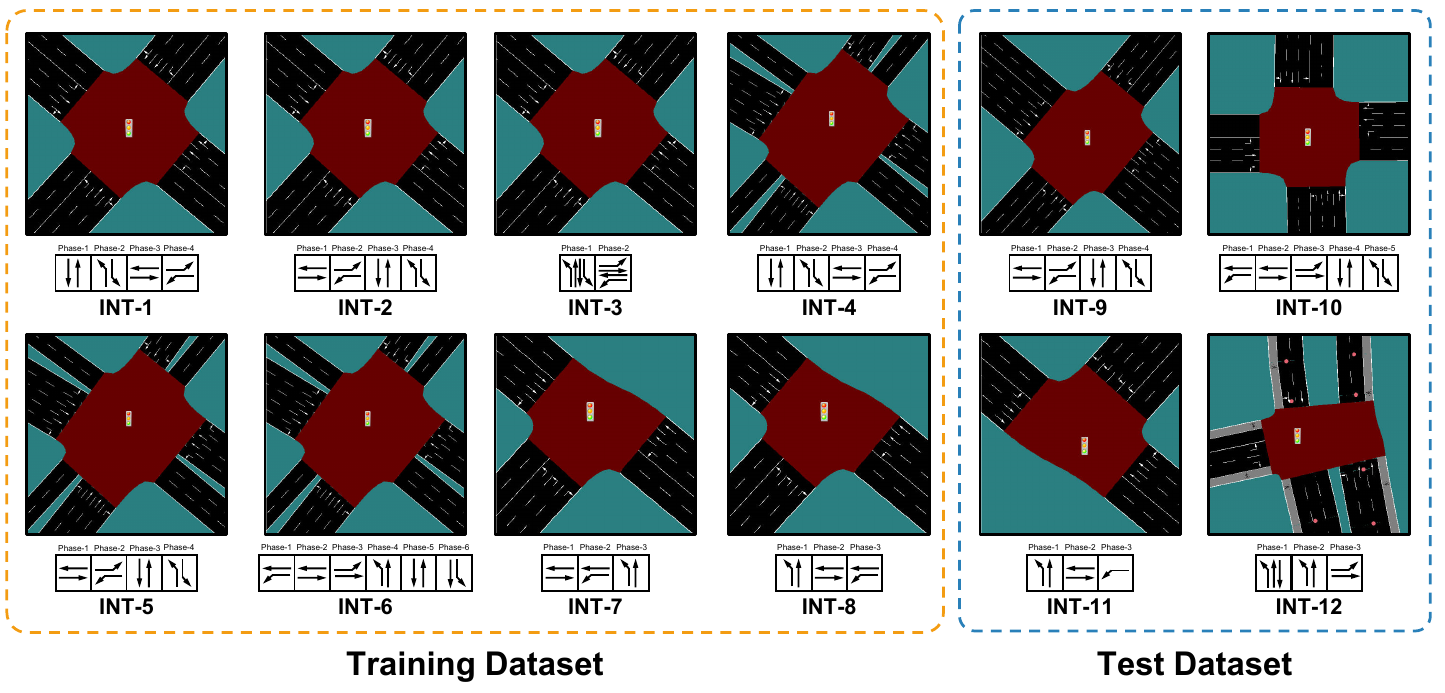}
    \caption{
        All intersection topologies with their available phases which used for the training and testing.
    }
    \label{fig_dataset}
\end{figure*}

Let us consider a pretrained weight matrix ${\bm W} \in \mathbb{R}^{n \times m}$ in the network accompanied by a LoRA module $\Delta {\bm W} = {\bm W}_{A} {\bm W}_{B}^{T}$, where ${\bm W}_{A} \in \mathbb{R}^{n \times d}$, ${\bm W}_{B} \in \mathbb{R}^{m \times d}$ with $d \ll n$. The output of this layer can be obtained as:

\begin{equation} \label{eq_lora}
    z = {\bm W} x + \Delta {\bm W} x = {\bm W} x + \frac{\alpha}{r} {\bm W}_{A} {\bm W}_{B}^{T} x,
\end{equation}
where ${\bm W}_A$ and ${\bm W}_B$ are initialized as a zero matrix and a zero-mean Gaussian distribution matrix, respectively. Furthermore, $\alpha$ is a constant scale hyperparameter whereas $r$ is the rank of the LoRA module. Through the combination of RL training using PPO and fine-tuning with LoRA, the proposed universal model can effectively address the challenges posed by important intersections of varying structures.

\begin{table*}[!htbp]
	\caption{Hyperparameter Setting}
	\label{tab_hparameters}
	\centering
        \begin{tabular}{|c|c|}\hline
            {\bf Hyper-parameter} & {\bf Value} \\\hline
            Learning rate & $0.0001$ \\\hline
            Trajectory memory size & $3000$ \\\hline
            Clipping range $\epsilon$ & $0.2$ \\\hline
            Discount factor $\gamma$ & $0.99$ \\\hline
            Value function coefficient $\lambda$ & $0.9$ \\\hline
            Scale hyperparameter $\alpha$ & $1$ \\\hline
            Rank of LoRA module & $8$ \\\hline
        \end{tabular}
\end{table*}

\section{Experiments} \label{sec_experiments}

\subsection{Experiment Settings}

Extensive experiments were conducted to validate the proposed model using the SUMO software package \cite{lopez2018microscopic} in this section. SUMO is an open-source microscopic traffic simulation tool designed for handling large networks. It provides the Traffic Control Interface (TraCI) for controlling traffic lights and retrieving traffic condition information for intersections. We calculate the flow and occupancy of each movement by analyzing the positions and trajectories of vehicles on the road. It is important to note that, in order to simulate real-world conditions, we consider only vehicles within proximity of $150$~m to the intersection, in lieu of all vehicles along the entire road. Moreover, 
a green light was followed by a yellow light of $3$~s before transitioning to a red light to ensure driver safety. The waiting time per vehicle was used as a performance metric to evaluate the effectiveness of the different methods. A low waiting time indicates that vehicles spent less time passing through the intersection. 

We utilized the Proximal Policy Optimization (PPO) implementation provided by the Stable Baselines3 library \cite{raffin2021stable}. To accelerate training, we employed $30$ parallel processes, and the total number of training environment steps was set to $10$M. The state representation included the previous $K=8$ snapshots of the junction matrix. The interval between two consecutive actions was $5$~s. 

The hyper-parameters are configured as shown in Table~\ref{tab_hparameters}. Furthermore, the actor and critic networks were designed as two-layer fully connected networks. The input sizes were $\{64, 32\}$, and the output sizes were $\{32, 2\}$ and $\{32, 1\}$ respectively for the actor and critic networks.

\begin{table*}[!htbp]
	\caption{All intersection configurations}
	\label{tab_dataset}
	\centering
	\resizebox{\textwidth}{!}{
		\begin{tabular}{cccccccccccccc}
			\hline
			& \multicolumn{8}{c}{\textbf{Training Dataset}} &  & \multicolumn{4}{c}{\textbf{Test Dataset}} \\ \cline{2-9} \cline{11-14} 
			Intersection ID & INT-1 & INT-2 & INT-3 & INT-4 & INT-5 & INT-6 & INT-7 & INT-8 &  & INT-9 & INT-10 & INT-11 & INT-12 \\ \hline
			roads & 4 & 4 & 4 & 4 & 4 & 4 & 3 & 3 &  & 4 & 4 & 3 & 3 \\
			lanes per road & (3,3,3,3) & (3,3,3,3) & (3,3,3,3) & (3,4,4,5) & (3,4,4,5) & (3,4,4,5) & (3,3,3) & (3,3,3) &  & (3,4,3,4) & (3,3,3,3) & (4,3,3) & (2,3,2) \\
			phases & 4 & 4 & 2 & 4 & 4 & 6 & 3 & 3 &  & 4 & 5 & 3 & 3 \\ \hline
	\end{tabular}}
\end{table*}

\subsection{Datasets} \label{sec_dataset}

This study considers intersections with diverse structures, aiming to employ one single universal model to predict actions for all intersections. Specifically, $12$ intersections of varying numbers of phases, lanes on each road, and approaching roads (i.e., $3$-way or $4$-way intersections) are constructed and used for experiments. Among these $12$ intersections, eight are used for training, while the remaining four are reserved for testing.  The topologies and phases of the $12$ intersections are depicted in Fig.~\ref{fig_dataset}, and Table~\ref{tab_dataset} provides a summary on their configurations. For instance, INT-4 shown in Fig.~\ref{fig_dataset} consists of four bi-directional approaching roads with three lanes in the north-south direction, five lanes in the west-east direction, and four lanes in each of the other two directions. INT-4 includes four phases, each combining two different movement signals. Consequently, the configurations for INT-4 in Table~\ref{tab_dataset} indicate the presence of four roads, lanes per road specified as $(3, 4, 4, 5)$ in clockwise order, and four phases.

There are three different intersection topologies in the training dataset. INT-1, INT-2 and INT-3 represent a regular $4$-way intersection scenario, with each road consisting of three lanes; INT-4, INT-5 and INT-6 represent a large $4$-way intersection scenario, featuring more than four lanes per road; INT-7, INT-8 and INT-9 depict the $3$-way intersection scenario. For the intersections with the same topology, we generate new intersections by altering the sequence of phases or the number of phases. For instance, INT-1 and INT-2 have identical configurations, but the sequence of phases differs. Similarly, INT-1 and INT-3 vary in terms of the number of phases. 

To assess the performance of the proposed model on unseen intersections, four testing scenarios are formed, namely INT-9, INT-10, INT-11 and INT-12. For instance, INT-9 and INT-10 also represent $4$-way intersections, but INT-9 features different lane configurations compared to the training set, while INT-10 differs in the number and combination of phases from the intersections in the training set. INT-11 modifies the lane and phase of each road based on INT-8. Finally, INT-12 simulates real-world traffic within the city of Ingolstadt, Germany \cite{ault2021reinforcement}. 

In addition to considering intersections with different structures, we generate $100$ unique pieces of the route for each intersection. Three-quarters of these routes were utilized for training, while the remaining quarter was reserved for evaluation. Each route has a duration of $30,000$ seconds, equivalent to approximately $8$ hours.

\subsection{Compared Methods}

To evaluate the performance of the proposed UniTSA, we compare the resulting universal model with several classic and state-of-the-art RL-based methods for TSC. 

\textbf{FixTime} \cite{miller1963settings}: Fixed-time control utilizes a predetermined cycle and phase duration plan, which is widely used in situations with steady traffic flow. We consider two versions of FixTime, FixTime-30 (Fix-30) and FixTime-40 (Fix-40). These variants correspond to fixed-time signal control plans where each phase has a duration of $30$ seconds and $40$ seconds, respectively.

\textbf{Webster} \cite{urbanik2015signal}: The Webster method determines the cycle length and phase split based on traffic volume during a specific period. It has been proven that when the traffic is uniform, the Webster method minimizes the travel time of all vehicles passing the intersection or maximizes the intersection capacity \cite{wei2019survey}. Additionally, the Webster method can be adapted for real-time applications. For fairness, we employ the Webster method to adjust the traffic lights based on real-time traffic in this study.

\textbf{SOTL} \cite{gershenson2004self}: Self-Organizing Traffic Light Control (SOTL) is an actuated signal control method that dynamically adjusts signal durations based on a manually determined threshold for the number of waiting vehicles. In this experiment, we set the threshold based on \cite{cools2013self}. 

\textbf{MPLight} \cite{chen2020toward}: MPLight incorporates the FRAP structure proposed in \cite{zheng2019learning} and a pressure-based reward mechanism inspired by \cite{wei2019presslight}. Furthermore, MPLight employs a sharing-parameter multilayer perceptron (MLP) to enhance adaptability across different intersection configurations. 

\textbf{AttendLight} \cite{oroojlooy2020attendlight}: This method adopts the attention mechanism to train a universal model capable of handling intersections with diverse structures and traffic flow distributions. It employs two attention models: the first attention model addresses variations in the number of roads and lanes, while the second attention model enables decision-making across intersections with different numbers of phases.

In addition to the baseline methods, we also consider several variations of our method:

\textbf{UniTSA (Single)}: This method focuses on training the model within a single environment. It utilizes the agent design described in Section~\ref{sec_agent_design} and employs an RNN-based structure for extracting information from the traffic state.

\textbf{UniTSA (Multi)}: In contrast to UniTSA (Single), this method trains the model across multiple environments simultaneously. We explore the performance of UniTSA (Multi) using various neural network designs, as discussed in Section~\ref{sec_feature_extraction}, including UniTSA (Multi+CNN), UniTSA (Multi+RNN), and UniTSA (Multi+Trans).

\textbf{UniTSA (Multi+TSA)}: This variant of UniTSA enhances the training process by incorporating five traffic state augmentation methods into UniTSA (Multi). It results in UniTSA (Multi+CNN+TSA), UniTSA (Multi+RNN+TSA), and UniTSA (Multi+Trans+TSA).

\begin{table*}[!htbp]
    \caption{Quantitative results (average waiting time per vehicle) of training intersections for universal models. A lower value indicates better performance and the lowest values are highlighted in bold.}
    \label{tab_results_in_training}
    \centering
    \begin{tabular}{lcccccccc}
    \hline
     & INT-1 & INT-2 & INT-3 & INT-4 & INT-5 & INT-6 & INT-7 & INT-8 \\ 
    \hline
    Fix-30 \cite{miller1963settings} & 39.458 & 38.862 & 8.323 & 35.123 & 35.031 & 51.923 & 18.920 & 18.595 \\
    Fix-40 \cite{miller1963settings} & 50.696 & 52.108 & 10.667 & 44.888 & 45.540 & 61.743 & 23.411 & 24.759 \\
    Webster \cite{urbanik2015signal} & 26.466 & 26.889 & 5.751 & 25.413 & 24.541 & 40.128 & 12.755 & 13.574 \\
    SOTL \cite{gershenson2004self} & 16.048 & 16.561 & \textbf{2.764} & 28.169 & 27.902 & 27.404 & 8.639 & 7.925 \\
    MPLight \cite{chen2020toward} & 19.111 & 14.659 & 4.469 & 16.067 & 19.925 & 19.115 & 6.654 & 7.523 \\ 
    AttendLight \cite{oroojlooy2020attendlight} & 16.483 & 13.893 & 3.860 & 16.903 & 18.915 & 20.795 & 6.532 & 8.104 \\
    \hline
    UniTSA (Multi+CNN) & 13.776 & 13.580 & 3.265 & 14.790 & 15.437 & 18.751 & 6.592 & 6.894 \\
    UniTSA (Multi+RNN) & 13.692 & 13.437 & 3.495 & 15.198 & 14.896 & 18.612 & 6.393 & 6.625 \\
    UniTSA (Multi+Trans) & 14.976 & 20.459 & 2.811 & 16.489 & 16.481 & 23.793 & 7.552 & 7.175 \\
    UniTSA (Multi+CNN+TSA) & 14.071 & 14.150 & 3.036 & 15.990 & 16.472 & 24.383 & 6.329 & 6.328 \\
    UniTSA (Multi+RNN+TSA) & 13.450 & 13.578 & 3.007 & \textbf{14.470} & \textbf{14.462} & 18.689 & \textbf{6.242} & \textbf{6.164} \\
    UniTSA (Multi+Trans+TSA) & \textbf{13.335} & \textbf{13.314} & 3.311 & 14.648 & 14.566 & \textbf{18.579} & 6.643 & 6.571 \\ 
    \hline
    \end{tabular}
\end{table*}

\subsection{Results of the training intersections}

In this section, the performance of the proposed UniTSA is compared against that derived from several existing approaches, including the FixTime approach, Webster model, SOTL, MPLight, and AttendLight, on the training intersections. In particular, UniTSA models of different network structures were examined. 

Table~\ref{tab_results_in_training} shows the average waiting time per vehicle achieved by UniTSA and other baseline methods on the training intersections. It is clear that the RL-based approaches demonstrated superior performance as compared to the conventional approaches in most intersection scenarios. Despite that SOTL achieved the shortest waiting time in the INT-2, it required manually defined thresholds for different environments, limiting its generalization in large-scale scenarios. 

Among the RL-based universal methods, namely MPLight, AttendLigh and UniTSA, UniTSA demonstrated significant performance improvements as compared to other RL-based methods. On average, UniTSA achieved $15\%$ and $12\%$ performance improvement over MPLight and AttendLight, respectively, across the eight intersections evaluated. When compared to MPLight, the proposed method incorporates not only parameter sharing techniques before replacing the MLP with RNN or Transformer block to capture the temporal information of the traffic state. In contrast to AttendLight, UniTSA simplifies the state design by representing the traffic state at a specific time instance using the junction matrix. Furthermore, five methods of traffic state enhancement are introduced, allowing the agent to observe a wider variety of traffic intersection states during training, which contributes to the improved performance of our model. 

Finally, we explored the impact of different network structures and traffic state augmentation. Inspection of Table~\ref{tab_results_in_training} reveals that an RNN-based structure yielded superior results compared to a CNN-based structure. Furthermore, in the {\em absence} of traffic state augmentation, UniTSA (Multi+CNN) and UniTSA (Multi+RNN) outperformed 
UniTSA (Multi+Trans) across most intersections, which is rather surprising. This can be attributed to the fact that the Transformer-based approach necessitates a larger volume of training data. However, in the {\em presence} of traffic state augmentation, the model can interact with a broader range of intersections with varying structures. As a result, UniTSA (Multi+Trans+TSA) outperformed UniTSA (Multi+RNN) and UniTSA (Multi+RNN+TSA) in many scenarios.

\begin{table}[!htbp]
    \caption{Quantitative results of test intersections for universal models.}
    \label{tab_results_in_test}
    \centering
    \begin{tabular}{lcccc}
    \hline
     & INT-9 & INT-10 & INT-11 & INT-12 \\ \hline
    Fix-30 & 40.341 & 38.360 & 17.136 & 17.440 \\
    Fix-40 & 60.043 & 50.580 & 23.504 & 20.570 \\
    Webster & 26.191 & 27.766 & 11.981 & 13.280 \\
    SOTL & 23.031 & 28.066 & 7.452 & 8.070 \\
    MPLight & 23.674 & 21.475 & 8.447 & 15.047 \\
    AttendLight & 18.080 & 18.501 & 7.393 & 12.982 \\ \hline
    UniTSA (Multi+CNN) & 17.877 & 18.244 & 7.063 & 12.820 \\
    UniTSA (Multi+RNN) & 16.771 & 15.450 & 6.807 & 10.140 \\
    UniTSA (Multi+Trans) & 17.640 & 16.269 & 6.598 & 9.630 \\
    UniTSA (Multi+CNN+TSA) & 13.075 & 14.245 & 6.794 & 7.550 \\
    UniTSA (Multi+RNN+TSA) & \textbf{12.677} & \textbf{14.208} & \textbf{5.914} & 6.730 \\
    UniTSA (Multi+Trans+TSA) & 13.054 & 16.186 & 5.983 & \textbf{6.190} \\ \hline
    \end{tabular}
\end{table}

\subsection{Results of the test intersections}

Next, we evaluate the key feature of UniTSA to see whether it can be utilized for {\em unseen} intersections. Four intersections, namely INT-9, INT-10, INT-11 and INT-12, are specifically used for the testing purposes. As discussed in Section~\ref{sec_dataset}, these intersections differ from the scenarios in the training set in terms of the number of lanes or the number of phases. Table~\ref{tab_results_in_test} summarizes the performance achieved by different methods, including baseline approaches and various variants of our proposed UniTSA method.  Consistent with the results obtained on the training set, the RL-based algorithms significantly outperformed the traditional traffic control algorithms. 

Among the RL-based universal methods, UniTSA models demonstrated superior performance across all test intersections as shown in Table~\ref{tab_results_in_test}. Among all the methods, UniTSA (Multi+RNN+TSA) and UniTSA (Multi+Trans+TSA) excelled in terms of reducing average travel time for vehicles passing through the intersections. UniTSA (Multi+RNN+TSA) showcases an average travel time reduction of approximately $32.9\%$ and $41.3\%$ as compared to MPLight and AttendLight, respectively. These results confirm the effectiveness of our approach in optimizing TSC and enhancing traffic flow efficiency.

Notably, incorporating traffic state augmentation techniques leads to improved performance among the different UniTSA variants. For instance, UniTSA (Multi+RNN+TSA), UniTSA (Multi+RNN+TSA), and UniTSA (Multi+RNN+TSA) exhibit enhancements of $23.4\%$, $19.8\%$, and $17.9\%$, respectively, when compared to UniTSA (Multi+RNN), UniTSA (Multi+RNN), and UniTSA (Multi+RNN). This improvement can be attributed to the inclusion of a greater variety of intersection scenarios within the training data through traffic state augmentation techniques, such as the ``Lane number change" method, which enables the generation of diverse combinations of lane configurations.

\begin{figure*}[!t]
    \centering
    \subfloat[]{\includegraphics[width=0.4\textwidth]{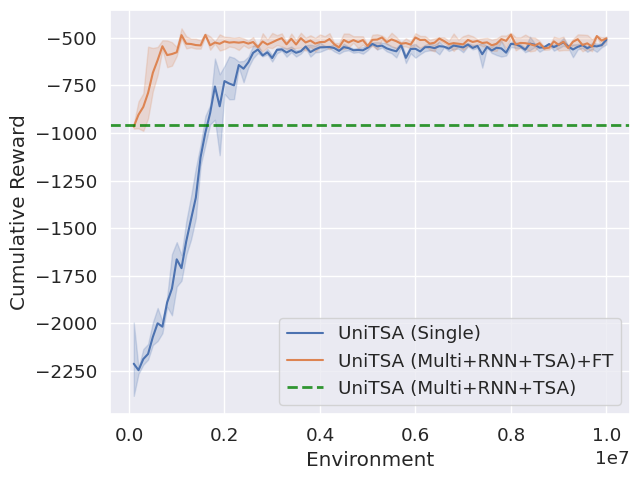}
    \label{fig_ft_1}}
    \hfil
    \subfloat[]{\includegraphics[width=0.4\textwidth]{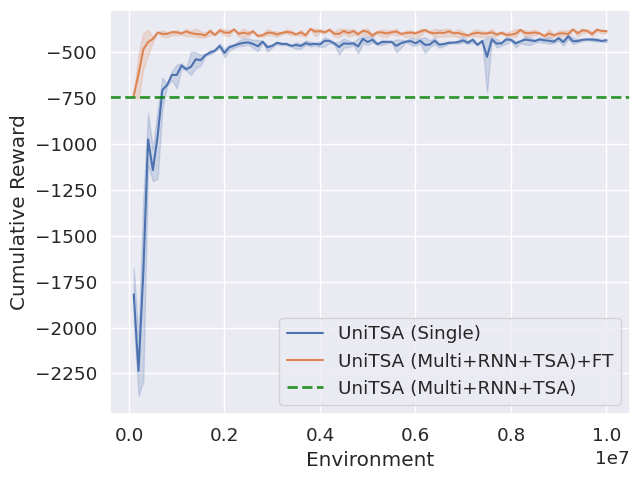}
    \label{fig_ft_2}}
    
    \hfill
    
    \subfloat[]{\includegraphics[width=0.4\textwidth]{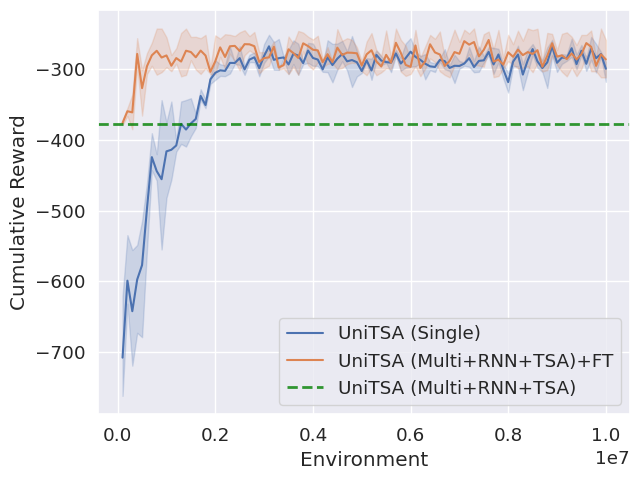}
    \label{fig_ft_3}}
    \hfil
    \subfloat[]{\includegraphics[width=0.4\textwidth]{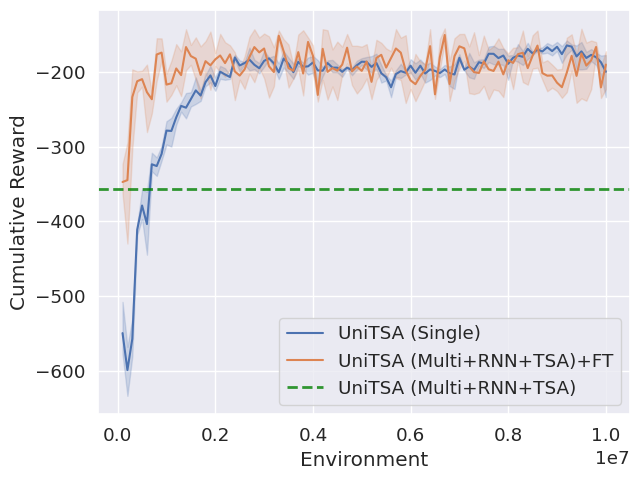}
    \label{fig_ft_4}}
    
    \caption{The environment steps of different methods at the four test intersections. (a) INT-9. (b) INT-10. (c) INT-11. (d) INT-12.}
    \label{fig_ft_analysis}
\end{figure*}

\begin{table}[!htbp]
    \caption{Quantitative results of fine-tuning in test intersections.}
    \label{tab_finetune_result}
    \centering
    \begin{tabular}{lcccc}
    \hline
     & INT-9 & INT-10 & INT-11 & INT-12 \\ \hline
    UniTSA (Multi+RNN) & 16.771 & 15.450 & 6.807 & 10.140 \\
    UniTSA (Multi+RNN+TSA) & 12.677 & 14.208 & 5.914 & 6.730 \\ \hline
    \multicolumn{5}{l}{\textit{1M Environment Steps}} \\ \hline
    UniTSA (Single) & 23.683 & 14.863 & 8.417 & 5.024 \\
    UniTSA (Multi+RNN+TSA) + FT & \textbf{10.995} & \textbf{12.553} & \textbf{4.475} & \textbf{3.534} \\ \hline
    \multicolumn{5}{l}{\textit{10M Environment Steps}} \\ \hline
    UniTSA (Single) & 10.353 & 12.254 & 4.359 & \textbf{3.089} \\
    UniTSA (Multi+RNN+TSA) + FT & \textbf{10.292} & \textbf{11.081} & \textbf{4.153} & 3.110 \\ \hline
    \end{tabular}
\end{table}

\subsection{Results of fine-tuning in test intersections}

In practical scenarios, certain intersections require special attention due to their significance. To address this, we begin with the universal model trained by UniTSA (Multi+RNN). The RNN-based UniTSA is chosen because it demonstrates superior performance across most intersections compared with both CNN-based and Transformer-based structures. In comparison to the UniTSA (Single), which is trained on a single scenario, the resulting model can quickly reach or even surpass the performance of the single-environment model after only a few training steps.

Fig.~\ref{fig_ft_analysis} shows the change of cumulative rewards over training steps for different models in the test intersections. The green dashed line represents the result of applying the universal model directly to new intersections without any fine-tuning. Notably, the model already exhibits promising results without any additional fine-tuning or transfer learning. The blue line represents the model trained from scratch whereas the orange line the fine-tuned model based on the universal model. It is observed that the single-environment model converges at approximately $3$M training steps. In sharp contrast, the fine-tuning model achieves comparable performance with only around $1$M training steps, resulting in approximately $66\%$ reduction in computation time while maintaining similar performance.

Table~\ref{tab_finetune_result} provides a detailed analysis of the model's performance after fine-tuning. At $1$M training steps, the fine-tuning model demonstrated an average performance improvement of $36\%$ as compared to UniTSA (Single) across the four test intersections. Even after $10$M training steps, the fine-tuning models continued to outperform the models trained from scratch by $3\%$. This aspect is particularly appealing for real-time applications. For instance, in a road network with over $1000$ junctions, it is possible to significantly reduce the number of interactions with the environment while maintaining comparable performance, thereby greatly enhancing training efficiency.

\begin{figure*}[!t]
    \centering
    \subfloat[]{\includegraphics[width=0.49\textwidth]{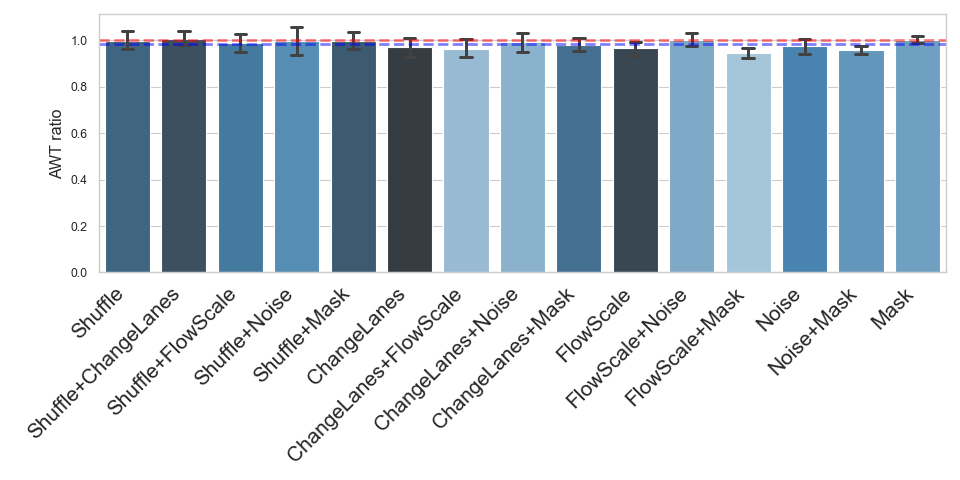}
    \label{fig_ad_analysis_1}}
    \hfil
    \subfloat[]{\includegraphics[width=0.49\textwidth]{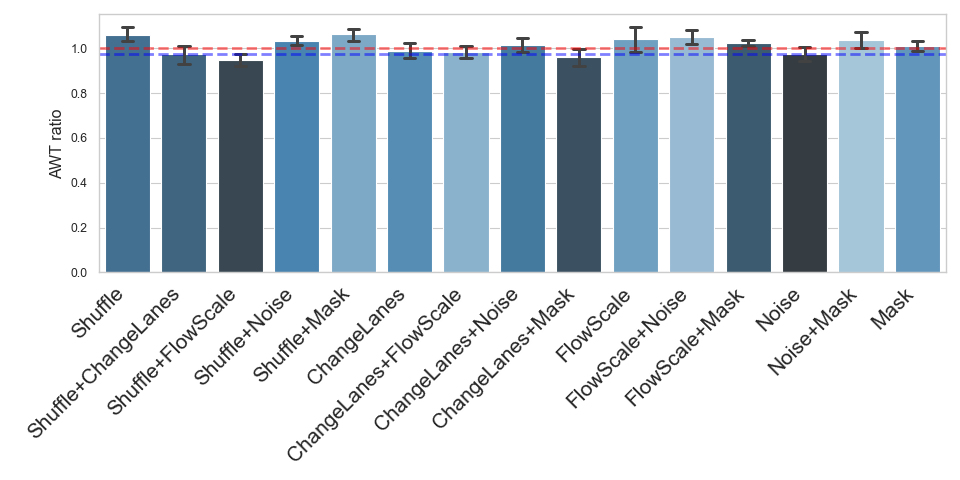}
    \label{fig_ad_analysis_2}}

    \hfill

    \subfloat[]{\includegraphics[width=0.49\textwidth]{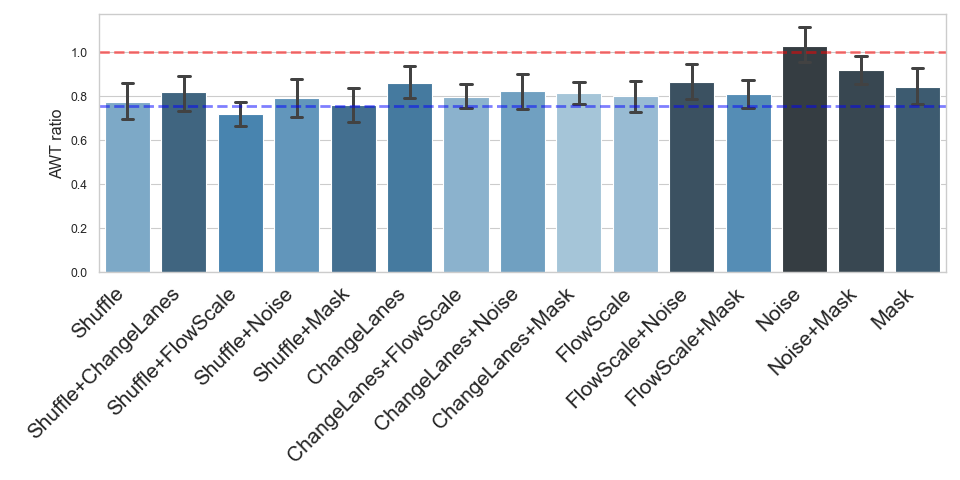}
    \label{fig_ad_analysis_3}}
    \hfil
    \subfloat[]{\includegraphics[width=0.49\textwidth]{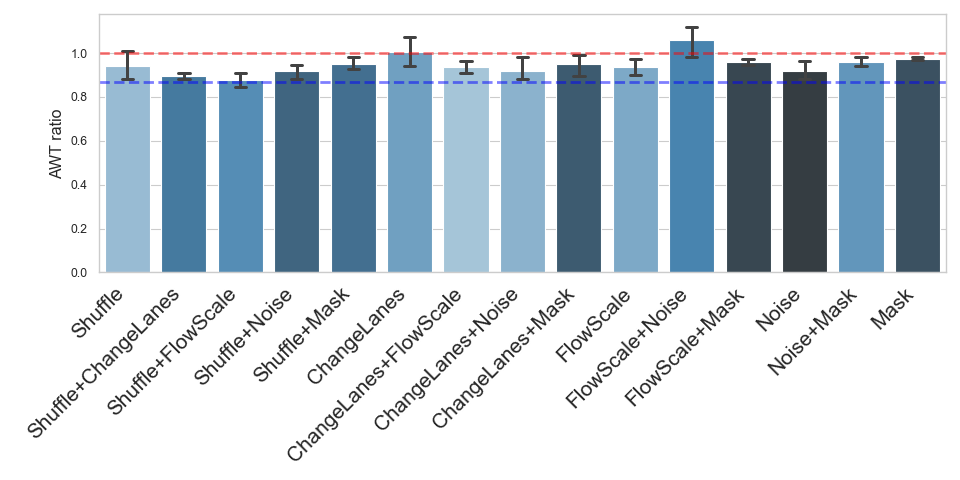}
    \label{fig_ad_analysis_4}}
    
    \caption{Comparative analysis of traffic state augmentation methods on the selected training and test intersections (a) INT-1. (b) INT-7. (c) INT-9. (d) INT-11.}
    \label{fig_ad_analysis}
\end{figure*}

\subsection{Comparative analysis on traffic state augmentations}

In this section, we analyze the effectiveness of traffic state augmentation methods in improving the performance of the UniTSA (Multi+RNN) model. We conducted experiments by applying different combinations of two traffic state augmentation techniques on both the training and test intersections. To evaluate the impact of these methods, we propose the Average Waiting Time (AWT) ratio between the models with and without traffic state augmentations with an AWT ratio of less than $1$ indicating an improvement achieved by the corresponding traffic state augmentation methods. 

Fig.~\ref{fig_ad_analysis} showcases the outcomes of the comparative analysis on INT-1, INT-7, INT-9, and INT-11 intersections, which represent common intersection structures encountered in real-world scenarios. Each bar in the figure corresponds to the average AWT ratio for a specific combination of traffic state augmentations, and the error bars represent the $95\%$ confidence interval. The blue dashed line represents the AWT ratio of UniTSA (Multi+RNN+TSA), which employs all available traffic state augmentation methods.

Fig.~\ref{fig_ad_analysis_1} and Fig.~\ref{fig_ad_analysis_2} depict the results obtained from the training intersections INT-1 and INT-7, respectively. Inspection of these figures reveals that the average performance improvement achieved through traffic state augmentations was around $2\%$. Furthermore, the inclusion of noise and mask methods incurred performance degradation in INT-7, which can be attributed to the fact that the model has already captured the underlying patterns and characteristics of the training intersections to a large extent. As a result, introducing additional variations through traffic state augmentation may not provide substantial benefits in the training set. However, traffic state augmentation methods demonstrated significant improvements when confronted with unseen intersection structures. 

Fig.~\ref{fig_ad_analysis_3} and Fig.~\ref{fig_ad_analysis_4} illustrate the AWT ratios for the test intersections INT-9 and INT-11, respectively. These results clearly demonstrate that most traffic state augmentation methods enhanced the performance of the base policy in the test intersections, which consists of unseen intersections. The diverse training samples generated through traffic state augmentation contribute to the improved performance. Among the traffic state augmentation techniques, movement shuffle and traffic flow scale were particularly effective in enhancing the model's performance. These techniques enable the model to adapt and learn from a wider range of scenarios, resulting in improved performance on the test set.

\section{Conclusion} \label{sec_conclusion}

In this paper, a universal RL-based TSC framework called UniTSA has been proposed for diverse intersection structures in V2X environments. More specifically, UniTSA offers the capability to train a universal RL agent by incorporating a junction matrix to characterize intersection states. To handle unseen intersections, new traffic state augmentation methods have been proposed to enrich the agent's data collection, resulting in improved performance and generalization for unseen intersection configurations. As a result, UniTSA eliminates the necessity of extensive customization and redevelopment for each individual intersection while offering a simple, efficient, and open-sourced implementation, which makes UniTSA a valuable framework for future research in data-efficient and generalizable RL-based TSC methods within the field of V2X. Extensive experimental results have demonstrated that UniTSA achieved the shortest average waiting time across various intersection configurations, surpassing the performance of the existing methods and outperforming the models trained from scratch with fine-tuning.

\section*{Acknowledgments}
This work was supported National Key Research and Development Program of China under Grant No. 2020YFB1807700 and the Shanghai Pujiang Program under Grant No. 21PJD092.

\bibliographystyle{unsrt}  
\bibliography{references}

\end{document}